

\input phyzzx

%
\catcode`\@=11 
\def\papers{\papersize\headline=\paperheadline\footline=\paperfootline}
\def\papersize{\hsize=40pc \vsize=53pc \hoffset=0pc \voffset=1pc
   \advance\hoffset by\HOFFSET \advance\voffset by\VOFFSET
   \pagebottomfiller=0pc
   \skip\footins=\bigskipamount \normalspace }
\catcode`\@=12 
\papers
\def\to{\rightarrow}

\tolerance=500000
\overfullrule=0pt

\pubnum={PUPT-1414 \cr
hep-th@xxx/930xxxx}

\date={July 1993}
\pubtype={}
\titlepage
\title{Fixed Point Structure For 3-D Rigid String}

\author{{
Weidong Zhao}\foot{Work supported by the NSERC postdoctoral scholarship
\nextline
e-mail: wdz@puhep1.princeton.edu
}}
\address{\it Joseph Henry Laboratories\break
Princeton University\break
Princeton, NJ
08544, USA}

\vskip 3.mm
\abstract{We show  that the usual fixed point for 3-d rigid string with
topological term appears to be a trivial one, consisting of  two decoupled
conformal
field theories. We further argue that by involving an additional
term  allowed by symmetries and thus generated by RG, the theory appears to
exhibit a new
fixed point with expected symmetry. The new fixed point is studied in the
weak- and strong coupling limit.

}

\endpage
\pagenumber=1

 \def\PL #1 #2 #3 {Phys.~Lett.~{\bf #1} (#2) #3}
 \def\NP #1 #2 #3 {Nucl.~Phys.~{\bf #1} (#2) #3}
 \def\PR #1 #2 #3 {Phys.~Rev.~{\bf #1} (#2) #3}
 \def\PRL #1 #2 #3 {Phys.~Rev.~Lett.~{\bf #1} (#2) #3}
 \def\CMP #1 #2 #3 {Comm.~Math.~Phys.~{\bf #1} (#2) #3}
 \def\IJMP #1 #2 #3 {Int.~J.~Mod.~Phys.~{\bf #1} (#2) #3}
 \def\JETP #1 #2 #3 {Sov.~Phys.~JETP.~{\bf #1} (#2) #3}
 \def\PRS #1 #2 #3 {Proc.~Roy.~Soc.~{\bf #1} (#2) #3}
 \def\IM #1 #2 #3 {Inv.~Math.~{\bf #1} (#2) #3}
 \def\JFA #1 #2 #3 {J.~Funkt.~Anal.~{\bf #1} (#2) #3}
 \def\LMP #1 #2 #3 {Lett.~Math.~Phys.~{\bf #1} (#2) #3}
 \def\IJMP #1 #2 #3 {Int.~J.~Mod.~Phys.~{\bf #1} (#2) #3}
 \def\FAA #1 #2 #3 {Funct.~Anal.~Appl.~{\bf #1} (#2) #3}
 \def\AP #1 #2 #3 {Ann.~Phys.~{\bf #1} (#2) #3}

\def \ins{\int d^2\sigma\sqrt{g}}

\def \pd{\partial}

\def \al{\alpha}
\def \be{\beta}
\def \eps{\epsilon}

One of the issues important in  current studies of
string theory is to understand the $d>1$ strings. A candidate for such
string theory is the
 rigid
string model proposed  by Polyakov\REF\POLa{A.M.Polyakov, \NP B268 1986 406 }
[\POLa]. In
this model, the string world-sheet action contains, in addition to the
usual Nambu-Goto area term, an extrinsic curvature
term which is designed to suppress the surfaces of high extrinsic curvature.
Such suppression is crucial  to  the formation of
a smooth phase,
because it eliminates the undesirable
dominance of those spiky surfaces in the configuration space, which
world have  caused problems in
defining a continuum string limit\REF\AMB{B.Durhuus, J.Froelich and T.Jonsson,
\NP B225 1983 185; \NP B240 1984 453}\REF\WAL{D.J.Wallace and R.K.P.Zia, \PRL
43 1979 308}\REF\DAV{F.David, \PL B102 1981 193; \NP B257 1985 543}
[\AMB-\DAV].

While this mechanism of suppression apparently works at the classical level,
a question remains as to whether it will still be
effective once the quantum
fluctuations are taken into account. Indeed, a renormalization group analysis
already showed that the extrinsic curvature is asymptotic free in the weak
coupling region[\POLa].
 It is thus possible that this term
might not survive
at large scale limit. This world be the case if there exists
 no  infra-red stable
fixed point at a
finite value of the coupling constant.  Of course, the existence of such fixed
point   cannot be determined by a  perturbative calculation alone,
and
 nonperturbative methods must be used. Unfortunately, in the meantime
we do not know of any effective means  to  systematically carry out
 such non-perturbative analysis.
As a result, it is not known in general if such fixed points will always
exist.

The situation  becomes a little simplier
in the $3$ dimensional spacetime. In this case, there are speculations that
an infra-red fixed point may exist, provided that a topological term with
half-integer coefficient is added to the action\REF\POLb{A.Polyakov, Princeton
Preprint, PUPT-1394(hep-th@xxx/93  (Apr. 1993) }[\POLb].
This point can be made
 most transparent
in the normal vector representation of the extrinsic curvature.
 This is the representation in which the extrinsic curvature
is expressed as the  kinetic term of the usual $O(3)$ non-linear $\sigma$
model,
${1\over \alpha}I_{1}={1\over 2\alpha}\ins\pd_{\alpha}n^i\pd^{\alpha}n_i$,
with  constraints that the vector $n^i$ being orthogonal to
the Riemann surface in  target space, whose dynamics is governed by the usual
gaussian action ${1\over \gamma}I_2={1\over 2\gamma}\ins
 \pd_{\alpha}x^i\pd^{\alpha}x^i$. The
topological term is $\theta I_{3}=
{i\theta \over 8\pi}\int d\sigma^2
\eps_{ijk}\eps^{\alpha\beta}n^i\pd_{\alpha}n^j
\pd_{\beta}n^k$,  representing the  winding number for $n$. It is well
known that, in the usual $O(3)$ model context, the presence of
 such a topological
term with a coefficient $\theta=\pi$
leads to a nontrivial fixed point, at which the model is described by the
$su(2)$
WZW model at the level $1$\REF\AFL{See, for example, I.Affleck, Les Houches
 Lecture Notes, 1988  in {\it Fields, Strings and Critical phenomena},
ed. E.Brezin and J.Zinn-Justin (North Holland)}[\AFL].
 Since the major
difference between the extrinsic
curvature  and the  $O(3)$ model is the orthogonality constraints
imposed on the vector $n^i$, one is naturally
led to speculate that a similar fixed point
for extrinsic curvature might also exist. Our first aim of this paper is to
show that this is indeed true. However, the fixed point turns out to be a
rather trivial one, in the sense that will become clear later.

Naively, the orthogonality constraints can be  implemented  by introducing
the
 lagrangian
multipliers $\eta^{\alpha}$. In the present case, due to  quantum fluctuations
these
 constraints  do not
survive at the infra-red limit. Indeed, it was shown [\POLb] that the induced
action
for $\eta^{\alpha}$
 possesses a logarithmic divergence,
so that a counterterm quadratic in $\eta^{\alpha}$ is required to ensure
the renormalizability. As a result,
the originally strict constraints are now
replaced by the smeared constraints,
taking the form $\lambda I_{4}={\lambda\over 2}\ins (n^i\pd_{\alpha}x^i)^2$
with $\lambda$
to be determined by the renormalization group flow. The total action thus can
be
written as
$$\eqalign{I&={1\over \alpha}I_1+{1\over \gamma}I_2+\theta I_{3}
+\lambda I_{4}\cr
&=\ins\left({1\over 2\alpha}(\pd_{\alpha} n^i)^2+{1\over 2\gamma}(\pd_{\alpha}
 x^i)^2+{\lambda\over 2}
(n^i\pd_{\alpha} x^i)^2\right)+{i\over 8} \int d^2\sigma
\eps_{ijk}\eps^{\alpha\beta}n^i\pd_{\alpha}n^j
\pd_{\beta}n^k.} \eqno(1)$$
We note that the metric used in this action is the induced metric, defined by
$g_{\alpha\beta}=\pd_{\alpha}x^i\pd_{\beta}x^i$.

As usual, the induced metric brings about
 difficulties in quantization.
We would like
to replace it by the intrinsic metric, introduced as independent variables.
The resulting  model  may or may not be
 equivalent to the original one, depending on whether in the infra-red limit
they
fall into the same universality class. In fact, using
the equation of motion one easily shows that $g_{\alpha\beta}= c_1
\pd_{\alpha}x\pd_{\beta}x+ c_2 \pd_{\alpha}n\pd_{\beta}n$ with some constants
$c_1, c_2$. It has been argued
\REF\POLc{A.Polyakov, private communication} [\POLc] that
the second term corresponds to a contribution from  higher derivatives, thus
may be regarded as being
irrelevant in the infra-red limit.  In this paper, we shall not
make any further discussion about  this point, and simply take
the action (1) with the intrinsic metric as our starting point.

We are now in a position to display the existence of the
fixed point. First of all, let us
assume that the world-sheet geometry is  flat.
We shall  follow closely the line of arguments presented
 in Ref \REF\SHAK{R.Shanker and N.Read, \NP B336 1990 457}[\SHAK], and thus
 work in the Hamiltonian formalism. For $\theta=\pi$,
 the
model in the strong coupling region can be mapped to an $1$-dimensional
half-spin chain [\SHAK], which is
 coupled to  $x$
through  $I_4$, with a gaussian
kinetic hamiltonian for $x$. The mapping is established through
 the relation
$n^i\to \sigma^i$, where $\sigma^i$ are the Pauli matrices associated with the
${1\over2}$ spins. Making use of this relation, the coupling term
can be shown to be  proportional
to $\sigma^i\sigma^j\pd x^i\pd x^j=\pd x^i\pd x^i$. This term has the same
structure
as the kinetic hamiltonian for $x$, thus  may be absorbed into the latter
 through a trivial renormalization.
In other words,
 the  degrees of freedom of  $x$  and
the $1$-d half-spin chain  are decoupled in the strong coupling limit.
The pure $1$-d half-spin chain is well
known to
exhibit a massless spectrum \REF\HAD{F.D.M.Haldane, \PL A93 1983 464}[\HAD].
Associated to this massless sector is an infra-red fixed point, to which
 any choice of bare coupling constant $\alpha$ will eventually flow.
 On general grounds, we expect that turning
on the gravity should not destroy the existence of such fixed point,
although its position and related exponents might be shifted.

This fixed point clearly
produces a $\alpha_c$ of finite  value, which is desirable for
the purpose of preserving the extrinsic
curvature term.  However, it appears to be
 a  trivial fixed point
 since the coupling between $n$  and  $x$ is to vanish
 at the infra-red limit.
This feature of
decoupling has already been seen in the strong coupling limit as discussed
above. We now argue that it is also true when the coupling constants
are close to the fixed point.
In this case, the model can be approximated by the $su(2)$
WZW model plus the usual gaussian action for $x$, with  perturbations
implemented by some (relevant or irrelevant) operators representing the
 deviation from the fixed point.
The crucial point is that, around this fixed point,
an operator coupling $n$ with $x$
must be
irrelevant. To see this,
 we observe  that in addition to $3$-d Poincare
symmetry the action (1) also
possesses a discrete symmetry $x^i\to -x^i$. The translational invariance of
$x$
implies
that any operator coupling to $x$
must depend on $x$ through $\pd_{\alpha} x^i$, whereas the discrete
symmetry requires that these $\pd_{\alpha}x^i$ appear in even power.
Obviously,  the
operator with the
lowest power in $\pd_{\alpha}x^i$
has the form
$A_{ij}(n)\pd_{\alpha}x^i\pd_{\beta}x^j$, with $A_{ij}$ being some primary
field of
the WZW model. Since $\pd_{\alpha}x^i\pd_{\beta}x^j$ already has a scaling
dimension $2$, in order to make the operator marginal we must require the
scaling dimension of $A_{ij}$  be zero. However, we know that in
the WZW model
the only primary field with zero conformal
dimension is the identity, which  corresponds to no coupling between
$n$ and $x$. It is thus clear that
 we are even unable to construct  a marginal operator
coupling  $x$ with $n$. For instance, since $n^i(z)n^j(0)\sim \delta^{ij}+
...$, we have $n^in^j\pd x^i\pd x^j \sim (\pd x)^2 + $ irrelevant terms.

We thus conclude that the fixed point  appears to describe
 a direct product of two conformal field theories with no coupling between
each other, and
the  orthogonality constraints have been completely smeared
 off. Although the $n$ field still survives at the infra-red limit, its
original nature of
being the normal vector to  the surface has  completely disappeared.
In other words,
in the large scale limit, there is still no
suppression to the surfaces of large extrinsic curvature. The triviality of
this fixed point is characterized by the
redundancy of the symmetry at the
conformal
point, $SU(2)_L\otimes SU(2)_R\otimes P(3)$\footnote{\dag}{There are
no independent
left- and right Poincare symmetries at the conformal point. This is contrary
to the
usual situation[\AFL].  I thank J.Distler
for pointing this out to me.}, whereas a proper fixed point would exhibit
a symmetry of $P(3)$ only.

Clearly, this perturbative analysis is valid
around
the particular fixed point
induced by the particular topological term $I_3$.
It does not, however, rule out the possibility for other
fixed points of different nature  which,
 if any, may be induced by  adding in the action terms other than this
topological term. In general, the terms to be added
should be renormalizable and respect to the symmetries of the microscopic
 theory, so that they may
be generated by renormalization.  We wish to find all of the terms that
satisfy this condition.

To this end, we
must first examine the  symmetries of the theory.
As already mentioned, the
Poincare group $P(3)$ and the discrete symmetry, $x^i\to -x^i$ are among
 the symmetry group of the action (1).
For $\theta=n\pi$ , the theory has
two extra discrete symmetries:  the world-sheet parity
$\sigma^1\to -\sigma^1$,
and the discrete symmetry $n^i\to -n^i$.
For a generic $\theta$, the last two operations are not
independent symmetries. However,
 their combination is still
a symmetry of the theory.

{}From physical grounds, this combination is necessary to the notion of
the target-space parity. In fact,
in 3-dimensional case the normal vector can be expressed
in terms of $x$  through the relation
$$n_i=\eps_{ijk}t^{jk}=
g^{-{1\over 2}}\eps^{\alpha\be}\eps_{ijk}\pd_{\alpha}x^j
\pd_{\be}x^k  \eqno(2)$$
where $t^{jk}$ is the antisymmetric tensor defined in
the $t$- representation of extrinsic curvature [\POLa]. A target-space parity
consists
of simultaneously flipping
the orientations of  $x^i$ and $n^i$, which is compatible with (2) only
 when
a world-sheet parity is  performed at the same time.
Since the target-space parity is always
a symmetry of the microscopic theory, it must be a  symmetry
 of the low energy effective theory.
This symmetry, along with the continuous $P(3)$ and the
discrete $x^i\to -x^i$, forms the symmetric group of the theory. Our question
is to find renormalizable terms other than $I_1,I_2,I_3,I_4$ which
respect to this symmetric group.

There is indeed a term, $I_5$, which meets this condition
$$I_5=\int d^2\sigma\eps_{ijk}n^i\pd_{\alpha}x^j
\pd_{\beta}x^k\eps^{\alpha\beta}. \eqno(3)$$
On general grounds, we
include it in the action with
a new dimensionless coupling constant
$\kappa$. The action now reads
$$I={1\over \alpha}I_1+{1\over \gamma}I_2+\theta I_{3}
+\lambda I_{4}+\kappa I_5. \eqno(4)$$
We wish to see whether this theory may flow to a
nontrivial fixed point at the infra-red limit.

In the rest of the paper, we shall provide evidence showing that this is very
likely to be true. We shall do so in both the weak- and strong coupling limits
in $\alpha$.
In the weak coupling limit, we will compute the $\be$- functions of the action
(4) at the $1$-loop level. The results will indicate that although  $\alpha$
is still asymptotic free, $\kappa$ is marginally relevant in
the infra-red limit. This produces a chance for a nontrivial
fixed point with nonvanishing coupling between $x$ and $n$. Moreover,
 we  will argue   in the strong coupling limit that  the
correlation function $<n^i(\sigma)n^j(0)>$ indeed decays algebraicly. This
usually implies that the theory is  on a critical line for $n$ fields for any
value of $\alpha$,
and it should eventually flow to a fixed point at which $n$ becomes massless.

We now begin to compute
the $\beta$ functions at the $1$-loop level. To do so,
  we make appropriate decompositions of  the fields, expand the
fast moving components
and keep
their quadratic terms\REF\POLd{A.M.Polyakov,
{\it Gauge fields and Strings} (Harwood Academic Press, New York, 1987)}
[\POLd]. This
$1$-loop
approximation generally
requires all the coupling constants to be small.  However, in the particular
 case
considered here,
since the
action is  quadratic  in $x$,  a correct $1$-loop expansion involves
summing up contributions to all orders in the coupling constants to $x$, which
are $\gamma, \lambda, \kappa$ in our case.
Consequently,
the $1$-loop
expansion becomes
a good approximation for any values of $\gamma, \lambda, \kappa$,
so long as $\alpha$ is  sufficiently small.

The easist way to perform such $1$-loop computation is to recast the theory
in the framework of the background non-linear $\sigma$ model, and making use of
 the standard formulae
for the background $\beta$ functions\REF\CAL{C.G.Callen, E.J.Martinec,
M.J,Perry and D.Friedan, \NP B262 1985 593}[\CAL]. To this end, we rewrite
the action (4)
as
$$I={1\over 2}\int d^2 \sigma \sqrt{g}g^{\al\be}\pd_{\al}x^{a}\pd_{\be}
x^{b}G_{ab}(x)+ \eps^{\al\be}\pd_{\al}x^a\pd_{\be}x^b B_{ab}(x)
\eqno(5)$$
In this equetion, both $n$ and $x$ are treated as background coordinates,
labeled by
$a,b=1,...,5$. In other words, we consider a 5-d manifold which is locally
a product of $S^2$ and $R^3$. Let $\theta^{\mu}, (\mu=1,2)$ be the
parametrization for $S^2$ (an often used example is
the spherical coordinates $\theta$ and
$\phi$), the background metric $G_{ab}$ and the antisymmetric tensor
$B_{ab}$  take the following block-diagonal form
$$ \matrix{G_{\mu\nu}={1\over \alpha}\pd_{\mu}n^i\pd_{\nu}n^i; &G_{ij}=
({1\over \gamma}\delta^{ij}+\lambda n^in^j); & G_{\mu i}=0 \cr
B_{\mu\nu}={\theta\over 8\pi}\eps_{ijk}n^i\pd_{\mu}n^j\pd_{\nu}n^k; &
B_{ij}=\kappa \eps_{ijk}n^k; & B_{\mu i}=0 \cr} \eqno(6)$$
where $n^i$ should always be understood as functions of $\theta^{\mu}$.
One may verify that
the action (4) can be recovered
by substituting $G_{ab}$ and $B_{ab}$ of (6) into (5).

We recall [\CAL] the formulae for the $1$-loop
$\be$ functions for $G_{ab}$ and $B_{ab}$. \footnote{\dag}{There
is no nontrivial dilaton background allowed by the symmetries.}
$$\eqalign{\beta^G_{ab}&={1\over 2\pi}\left(R_{ab}-
{1\over 4}H_{acd}H^{cd}_b\right) \cr
\beta^B_{ab}&={1\over 2\pi}\nabla^cH_{abc}}\eqno(7)$$
where $H_{abc}=3\pd_{[a}B_{bc]}$ is the strength for $B$.
Given $G_{ab}$ and $B_{ab}$ in (6), these $\beta$ functions
  can be calculated
straightforwardly. The resulting
 $\beta^G_{ab}$ and $
\beta^B_{ab}$ must have same
structures (with different coefficients)
 as $G_{ab}$ and $B_{ab}$, owing to the renormalizability.
 By comparing the coefficients, we obtain the following
$\beta$ functions for various couplings
$$\eqalign{\beta_{\al}&=-{1\over 2\pi}\alpha^2\left(1-{(\lambda^2+
\kappa^2)\gamma^2\over
2(1+\lambda\gamma)}\right) \ + \ \ {\cal O}(\alpha^3) \cr
\beta_{\gamma} &=-{1\over 2\pi}\alpha\left({(\lambda^2-\kappa^2)\gamma
\over 2(1+\lambda\gamma)}-2\lambda\right) \ + \ \ {\cal O}
(\al^2)\cr
\beta_{\lambda}&={1\over 2\pi}\alpha\left(3\lambda+{1+2\lambda\gamma\over
2(1+\lambda\gamma)}(\lambda^2-\kappa^2)\gamma \right) \ +
\ \ {\cal O} (\al^2)\cr
\beta_{\kappa}&=-{1\over 2\pi}\alpha{2\kappa\over 1+\lambda\gamma}\ + \ \
{\cal O} (\al^2)\cr
\beta_{\theta}&=0}\eqno(8) $$
These complicated expressions as functions of $\gamma, \lambda, \kappa$  are
a consequence of  summing up  contributions  to all orders in
 these couplings, a point we discussed earlier.

A few remarks are in order. First, at the perturbative level, the topological
term $I_3$ does not appear
to have any effect to the renormalization.
 In fact, $\beta_{\theta}=0$ to all orders in $\alpha$. It is thus consistent
to forget about it from now on.  Secondly, the unitarity of the theory
 requires
$\alpha$,  $\gamma$ and  $\lambda$   be
non-negative, whereas $\kappa$ be purely imaginary. Thirdly,
 by a trivial scaling of $x$ we observe
that out of the three $x$-related coupling constants $\gamma, \lambda, \kappa$
only two are independent. It is  convenient to introduce the
following  independent couplings
$u=\lambda\gamma$ and $v=i\kappa\gamma$. In terms of $u$
and $v$, the $\beta$ functions (8) can be reduced to
$$\eqalign{\beta_{\alpha}&=-{\alpha^2\over 2\pi}(1-{u^2-v^2\over 2(1+u}) \cr
\beta_u &={\alpha\over 4\pi}(3u^2+6u+v^2)\cr
\beta_v &={\alpha\over 2\pi}{ v\over 2(1+u)}(u^2+2u-4-v^2)}\eqno(9)$$
Note that the unitarity requires $\alpha$, $u$ be non-negative, whereas $v$
be real.
We emphasize that these equations are valid for small value of $\alpha$
and
 with no restriction on $u$ and
$v$ except the unitarity requirement.

The RG flow in the infra-red limit can be obtained  by examining
 the $\beta$ function equestions at $t=log\Lambda\to -\infty$
$$\eqalign{&{d\over d log\Lambda}\alpha =\be_{\al}; \cr
&{d\over d log\Lambda}u=\be_u; \cr
&{d\over d log\Lambda}v=\be_v. }\eqno(10)$$
Since we are unable to solve these  equestions analytically, we will only
 estimate the
asymptotic behaviours for the couplings.

We begin by observing that $\be_u$ is always positive for couplings
in the unitary region. Therefore $u$ should flow
to zero in the infra-red limit. For $u$ being sufficiently small
or  $v^2$ being sufficiently large
so that $u^2+2u < 4+v^2$,
$\be_{v}$ is dominated by  two  terms of negative sign. The negative sign
indicates that $v^2$ will
 flow to infinity at the infra-red limit. On the other hand, if the initial
$u$ and
$v$ are
such that $u^2+2u > 4+v^2$, $\beta_{v}$ appears to be positive and
thus produce a decreasing $v^2$  in the infra-red limit.
However, since $u$ will eventually flow  to zero, this decreasing trend of
$v^2$ is stopped at last when $u$ reaches the point $u^2+2u=4$, after which
the condition $u^2+2u < 4+v^2$ is satisfied  and $v^2$ increases again.
In other words, the
coupling $v$ will eventually flow to either infinities depending on its sign,
regardless of the initial value of $u$.

By a similar argument, one can show that $\be_{\al}$ in the infra-red limit
is also
dominated
by the contribution from $v$, which takes a negative sign indicating
an asymptotic freedom for $\alpha$. The
 overall infra-red behaviour for
$v$ and $\alpha$ is $\alpha \sim v \sim (log\Lambda)^{-{1\over 3}}$ for
$\Lambda \sim 0$. We
conclude that in the infra-red limit
 the couplings $\al$ and $v$ are marginally relevant, whereas the
coupling $u$ is marginally irrelevant.

This conclusion is of course incomplete, as $\alpha$ will eventually flow
 to a region where these $1$- loop results are no longer valid. Taking
into account the higher order contributions, there is a possibility
that $\al$ and $v$ may flow to a fixed point with some
finite values. If this is
indeed true, the non-vanishing $v$ should
 indicate a nontrivial coupling between
$n$ and $x$, corresponding to a single $P(3)$ as  symmetry.

Further evidence in favor of this scenario may be found
 by investigating the large-distance behaviour of certain correlation functions
at the strong coupling
limit. Usually,  an algebraicly decaying behaviour for such a correlation
function
is an indication that the  critical
line extends up to the strong coupling limit. This means that
   a strong coupling constant  flows
to a
fixed point of finite value with only massless sector left over. In our case,
the relevant
 correlation function is $<n^i(\sigma,t)n^i(0)>$, whose large-distance
behaviour  is determined
by those low-lying states  that can couple to $n$. If in the hamiltonian
there is no couplings
between $n$ and $x$, the low-lying spectrum contributing
to this correlation function is identical to
those of usual O(3) $\sigma$ model, despite the existence of
 the massless sector for $x$. Subsequently, $<n^i(\sigma,t)n^i(0)>$
decays exponentially for large $\sigma$, indicating a massive phase
characterized by $\alpha=\infty$. For this reason, it is clear that
in order to achieve a massless
phase  a coupling between $x$ and $n$ is needed. Such a role is exactly
played by the the coupling term $I_5$, as  will be shown below.

In general, the correlation function $<n^i(\sigma,t)n^i(0)>$
can be expanded as
$$ <n^i(\sigma,t)n^i(0)>=\sum_m {_{ph}<0|n^i(\sigma)}|m><m|n^j(0)|0>_{ph}
\exp^{-t(\eps_m-\eps_0)},
 \eqno(11)$$
with a complete set of eigenstates $\{|m>\}$ of hamiltonian. The
hamiltonian can be derived from the lagrangian (4) by the standard canonical
quantization.
 Since we
are  interested in the large $t$ behaviour, only the low-lying
spectrum contributes to (11). In the strong coupling limit, these low-lying
 states
 consists of tensor-products of the unperturbed  $l=0$
sector for $n$ and various  particle states for $x$, mixed with some higher
$l$ components for $n$ induced by the perturbation. Furthermore,
the symmetries of the theory give rise to
certain
selection rules which to the lowest order
only allow $l\le 1$ components to couple
to the form factor $ <m|n^i(\sigma)|0>_{ph}$. Restricted to this subspace,
 we find the following effective hamiltonian
$$H=c_1l(l+1)+
c\int_0^{\Lambda} dk |k|a^{\dag i}(k) a^{i}(k) + H_{int} \eqno(12)$$
where
$$H_{int}={c_2\over 2}\int^{\Lambda} dk_1dk_2F(k_1,k_2)
\eps_{ijk}n^k(-k_1-k_2)\left(a^{\dag i}(k_1)a^{\dag j}(k_2)-a^{\dag
i}(k_1)a^j(k_2) +c.c.
\right) \eqno(13)$$
$$F(k_1,k_2)=|k_1k_2|^{1\over 2}(sign(k_1)-sign(k_2)).$$
In deriving (12) and (13), we have dropped terms corresponding to
 $n^in^j\pd x^i\pd x^j$ and $(\pd_{\sigma}n^i)^2$, as to the lowest order
they do not couple to
the form factor  due to the violation of
the $n$-parity conservation.
Their contributions in higher orders have been absorbed into the renormalized
coupling constants $c, c_1, c_2$. The first term  in (12)
is the  quadratic casimir
arising from the kinetic term in $n$,
and to the lowest order we only consider  $l=0$ and $l=1$.
 As usual, we have expressed $x^i(\sigma), p^i(\sigma)$
through their
momentum oscillations $a^{\dag i}(k), a^i(k)$. In a similar fashion,
$n^i(\sigma)$ has also be expressed in the momentum space.
Note that $H_{int}$, with a coupling $c_2$ proportional to $v$,
 arises exactly from $I_5$.

The correlation function (11) can be easily calculated using the effective
hamiltonian (12). To the lowest order,
 the only states contributing
to the form factor are the  two particle states $|k_1,i;k_2,j>$
 of momentum
$k_1,k_2$ and polarization $i,j$, which can be expressed using the standard
perturbation theory as
$$|k_1,i;k_2,j>=a^{\dag i}(k_1)a^{\dag j}(k_2)|0>+c_2
{F(k_1,k_2)\eps^{lij}n^l(-k_1-k_2)\over 2c_1-c(|k_1|+|k_2|)}
|0> \ + {\cal O}(c_2) \eqno(14)$$
To the same order, the ground state $|0>_{ph}$ can be expressed as
$$|0>_{ph}=|0>+{c_2\over 2}\int^{\Lambda}
dk_1dk_2{F(k_1,k_2)\eps_{ijk}n^k(-k_1-k_2)
\over 2c_1+c(|k_1|+|k_2|)}
a^{\dag i}(k_1)a^{\dag j}(k_2)|0> \ + {\cal O}(c_2)  \eqno(15)$$
Substituting (14) and (15) into (11), we find
$$<n^i(0,t)n^i(0)>\sim \int^{\Lambda}
 dk_1dk_2 \left({c_2\over c_1}F(k_1,k_2)\right)^2
e^{-c(|k_1|+|k_2|)t} \eqno(16)$$
Thus $<n^i(0,t)n^i(0)>\sim t^{-4}$ for large $t$. It is easy to see that
higher order perturbations in $c_2$ are irrelevant in the large $t$ limit.

In conclusion, we have shown in this paper that the new term $I_5$, which is
consistent
with the symmetry, is marginally relevant in the
weak coupling limit. In the strong coupling limit, the correlation function
$<n^i(0,t)n^i(0)>$ decays algebraicly for large $t$. These are indications
for a massless phase for $n$. Furthermore, since the coupling constant $v$
is presumably nonzero at the fixed point, the relation (2) holds at the
conformal point, and
the symmetry of the low energy theory is $P(3)$, as desired. It world be
very interesting to explicitly find out this fixed point, as well as the
various exponents related. Clearly, a lot of works still need to be done.

\ack I am very grateful to
A.Polyakov for bringing
 my interests to this topic,
as well as numerous enlightening discussions. I world also
like to thank  J.Distler, M.Douglas, D.Gross, M.Li, A.A.Ludwig
 for  useful conversations.

\refout

\end